\documentclass[11pt]{article}
\usepackage{latexsym}
\usepackage{epsfig}

\newcommand{\be}{\begin{equation}}
\newcommand{\ee}{\end{equation}}
\newcommand{\ba}{\begin{eqnarray}}
\newcommand{\ea}{\end{eqnarray}}
\newcommand{\bi}{\begin{itemize}}
\newcommand{\ei}{\end{itemize}}
\newcommand{\baa}{\begin{array}}
\newcommand{\eaa}{\end{array}}

\newcommand{\nr}[1]{(\ref{#1})}

\newcommand{\xt}{\mathbf{x}_T}
\newcommand{\yt}{\mathbf{y}_T}
\newcommand{\rmi}[1]{{\mbox{\scriptsize #1}}}
\newcommand{\fr}[2]{{\frac{#1}{#2}\,}}
\newcommand{\fra}[2]{\textstyle{\frac{#1}{#2}\,}}  
\newcommand{\mn}{{\mu\nu}}
\newcommand{\bfx}{{\bf x}}

\newcommand{\gp}{g(x^+)}
\newcommand{\fm}{f(x^-)}

\newcommand{\eq}{Eq.~}

\def\CL{{\cal L}}

\def\gsim{\raise0.3ex\hbox{$>$\kern-0.75em\raise-1.1ex\hbox{$\sim$}}}
\def\lsim{\raise0.3ex\hbox{$<$\kern-0.75em\raise-1.1ex\hbox{$\sim$}}}

\begin{document}

\begin{titlepage}
\begin{flushright}
HIP-2007-75/TH\\
\end{flushright}
\begin{centering}
\vfill

{\Large{\bf Gravity dual of conformal matter collisions in 1+1 dimensions
}}

\vspace{0.8cm}

\renewcommand{\thefootnote}{\fnsymbol{footnote}}

K. Kajantie$^{\rm a}$\footnote{keijo.kajantie@helsinki.fi},
Jorma Louko$^{\rm b}$\footnote{jorma.louko@nottingham.ac.uk},
T. Tahkokallio$^{\rm a,c}$\footnote{touko.tahkokallio@helsinki.fi}

\setcounter{footnote}{0}

\vspace{0.8cm}

{\em $^{\rm a}$%
Department of Physics, P.O.Box 64, FI-00014 University of Helsinki,
Finland\\}
{\em $^{\rm b}$%
School of Mathematical Sciences, University of Nottingham,
Nottingham NG7 2RD, UK
\\}
{\em $^{\rm c}$%
Helsinki Institute of Physics, P.O.Box 64, FI-00014 University of
Helsinki, Finland\\}

\vspace*{0.8cm}

\end{centering}

\noindent
We find the three-dimensional gravity dual of a process in which two
clouds of (1+1)-dimensional conformal matter moving in opposite
directions collide. This gives the most general conformally invariant
holographic flow in the 1+1 dimensional boundary theory in terms of
two arbitrary functions.  With a suitable choice of the arbitrary
functions the process can be interpreted as an opaque collision of two
extended systems with central, fragmentation and interaction
regions. Comparison with classical gluon field calculations relates
the size of the system with the saturation scale.

\vfill \noindent


\vspace*{1cm}

\noindent

31 December 2007, revised 11 January 2008

\vfill

\end{titlepage}

\section{Introduction}
The final state of a heavy ion collision has been recently
studied in gauge/gravity duality
in a number of papers
\cite{jp,jp2,nakamura1,nakamura2, janik1,janik2,kt,kovchegov,klt,%
Lin:2006rf,Nakamura:2007nx,Hatta:2007cs,%
Benincasa:2007tp,Alsup:2007bs,Bhattacharyya:2007jc}.
The early stages of the collision process are more difficult to model and have so far
received less attention~\cite{ssz}.
In this paper, we consider both the final state and the onset of the collision in a
simplified model in which the boundary theory is (1+1) dimensional and the bulk is
locally AdS$_3$.
We shall give explicitly the general bulk solution that is dual to
two colliding clusters of conformal boundary matter.
That such a general solution
can be found is, of course, due to the simplicity of gravitational
dynamics in three dimensions. The solution involves two arbitrary
functions, each depending on only one of the light cone coordinates.
With a
suitable choice for these two functions, the boundary process can be described in
the language of heavy ion collisions as having two incident extended
systems, an interaction diamond and central and fragmentation regions
in the final state. There is one important difference between
our boundary matter and that of real (3+1)-dimensional heavy ion collisions though:
our matter is extremely opaque and bounces off backwards in the collision.

As a special case, our bulk solution also provides the metric of two
colliding shock waves in three dimensions. This metric describes a spacetime in
which two massless point particles create a three-dimensional black hole
\cite{matschull,holst}.
In four (or more) spacetime
dimensions the metric of a single shock wave is known both with a
vanishing \cite{as,dh} and nonvanishing
\cite{hottatanaka,podolskygriffiths,horoitzhaki} cosmological
constant, and an extension to colliding waves would have important
applications in the study of black hole production at
colliders~\cite{kaloperterning}, but finding such a solution has
proved a formidable task~\cite{death1,death2}.

We use the three-dimensional coordinates $(x^+,x^-,z)$,
$(\tau,\eta,z)$ and $(t,x,z)$, referred to respectively as the light
cone, Milne and Minkowski coordinates. The conformal boundary is at $z=0$,
the coordinates are at $z=0$ related to each other by
\be
x^\pm=
{t\pm x\over\sqrt2}=
{\tau\over\sqrt2} \, e^{\pm\eta},\qquad
t=\tau\cosh\eta,\,x=\tau\sinh\eta ,
\label{lcc}
\ee
where $\tau>0$, and the boundary Minkowski metric takes the form
\be
ds^2=-2dx^+dx^-= -d\tau^2+\tau^2d\eta^2=-dt^2+dx^2 .
\label{eq:1+1mink}
\ee

\section{Colliding conformal boundary matter}
\label{sec:genbulksolution}

In three dimensions, the metric of a plane wave moving in the $x^+$
direction can be written in the light cone coordinates as
\be
g_{MN}^+=\frac{\CL^2}{z^2}
\left( \begin{array}{ccc}
  0 & -1& 0 \\
  -1& z^2f(x^-) & 0 \\
  0 & 0 & 1
  \end{array} \right) ,
\label{gp}
\ee
where $f(x^-)$ is an arbitrary function of~$x^-$.
This metric satisfies the AdS$_3$ Einstein equations
\be
R_{MN}-\fra{1}{2}R\, g_{MN}-{1\over\CL^2}g_{MN}=0,
\label{einstein-equations}
\ee
following from the action
\be
S={1\over16\pi G_3}\int d^3x\sqrt{-g}\left(R+{2\over\CL^2}\right),
\label{einstein-action}
\ee
and must therefore be locally isometric to AdS$_3$.  In the special
case in which $f(x^-)$ is proportional to~$\delta(x^-)$, we obtain the
three-dimensional negative cosmological constant generalisation of the
Aichelburg-Sexl (AS) shock wave~\cite{as}.

By symmetry, the metric of a plane wave moving in the $x^-$ direction
can be written as
\be
g_{MN}^-=\frac{\CL^2}{z^2}
\left( \begin{array}{ccc}
 z^2g(x^+) & -1& 0 \\
  -1& 0 & 0 \\
  0 & 0 & 1
  \end{array} \right),
\label{gm}
\ee
where $g(x^+)$ is an arbitrary function. Note that in each case the
metric components are independent of the light cone coordinate in
whose direction the wave is travelling.

While extensions of the above plane wave solutions to four and more dimensions
are known \cite{hottatanaka,podolskygriffiths,horoitzhaki}, an extension that would
describe two colliding waves in four or more dimensions has proved elusive. In three
dimensions, however, we can find such a solution
by making in the light cone coordinates $(x^+,x^-,z)$ the ansatz
\be
g_{MN} =
\frac{\CL^2}{z^2}
\left( \begin{array}{cc}
  g_\mn & 0 \\
  0  & 1
  \end{array} \right),
\ee
where the two-dimensional metric components $g_\mn$ may a priori depend on any of the coordinates.
If $g_\mn$ is assumed to reduce at $z\to0$ to the (1+1)-dimensional Minkowski
metric~\nr{eq:1+1mink}, the general solution to Einstein's equations
\nr{einstein-equations} may be shown to be
\be
g_{MN}=\frac{\CL^2}{z^2}
\left( \begin{array}{ccc}
  z^2 g(x^+) & -1-{z^4\over 4}g(x^+)f(x^-)& 0 \\
  -1-{z^4\over 4}g(x^+)f(x^-)& z^2 f(x^-)& 0 \\
  0 & 0 & 1
  \end{array} \right),
\label{gensollc}
\ee
where $g$ and $f$ are arbitrary functions, each of
dimension $1/{(\textrm{distance})}^2$.
This solution can also be obtained by specialising that given in
\cite{ss} to a flat boundary metric.
It may be regarded as a superposition of
the plane wave \nr{gp} moving in the $x^+$ direction and the
plane wave \nr{gm} moving in the $x^-$ direction, the only nonlinearity in the superposition
being a correction in the component~$g_{+-}$.
The solution can thus be characterised as a collision of two plane waves.
If $f$ and $g$ are each proportional to the delta-function, the solution describes the
collision of two shock waves.

To analyze the bulk solution \nr{gensollc} from the boundary theory viewpoint,
we compute the energy-momentum tensor on the conformal boundary at $z=0$~\cite{skenderis}.
The small $z$ expansion of $g_\mn$ reads
\ba
&&g_\mn(x^\pm,z)=g_\mn^{(0)}+g_\mn^{(2)}z^2+ \ldots
\nonumber
\\[1ex]
&=&\left( \begin{array}{cc}
0& -1\\
-1 & 0
\end{array} \right) +
\left( \begin{array}{cc}
g(x^+) & 0\\
0 & f(x^-)
\end{array} \right)z^2 -
\left( \begin{array}{cc}
0 & \fr14g(x^+)f(x^-)\\
\fr14 g(x^+)f(x^-) & 0
\end{array} \right)z^4 ,
\nonumber\\
&&
\ea
from which we obtain
\be
T_{\mu\nu}=\frac{\CL}{8\pi G_3}
[g_{\mu\nu}^{(2)}-g_{\mu\nu}^{(0)}\textrm{Tr}(g_{\mu\nu}^{(2)})]
=\frac{\CL}{8\pi G_3}
g_{\mu\nu}^{(2)}=\frac{\CL}{8\pi G_3}
\left( \begin{array}{cc}
g(x^+) & 0\\
0 & f(x^-)
\end{array} \right).
\label{skende}
\ee
$T_{\mu\nu}$ is clearly conserved. A~conserved total energy $E$ can
hence be defined as the integral of $T^{tt}$ over
a surface of constant~$t$. From \nr{skende} we find that
$E$ decomposes as
\ba
E &=&
E^+ + E^- ,
\nonumber
\\
E^+
&=&
\frac{\CL}{8\pi G_3}{1\over\sqrt2}\int_{-\infty}^\infty dx^- \, \fm ,
\nonumber
\\
E^-
&=&
\frac{\CL}{8\pi G_3}{1\over\sqrt2}\int_{-\infty}^\infty dx^+ \, \gp,
\label{etotplusminus}
\ea
where $E^+$ and $E^-$ are respectively the energies of the components moving in the
direction of $x^+$
and~$x^-$.


When $g=0$ but $f\ne0$, the energy-momentum tensor can be written in
terms of the null vector $n^\mu=(1,0)$ as
\be
T_\mn=\frac{\CL}{8\pi G_3}\fm n_\mu n_\nu,\qquad n_\mu=(0,-1) ,
\label{dust}
\ee
which shows that the boundary matter is null dust moving in the $x^+$
direction.  Similarly, when $f=0$ but $g\ne0$, the boundary matter is
null dust moving in the $x^-$ direction. When both $f$ and $g$ are
nonvanishing, the energy-momentum tensor is the sum of
the left-moving and right-moving parts, and as $f$ and $g$ are
independent, it is possible to interpret the boundary matter as two null
dust clouds that just pass through each other without interacting.  In
higher dimensions interactions are however present, and the boundary
energy-momentum tensor after the collision is then usually interpreted
as a single-component fluid with internal dynamics and
thermodynamics~\cite{bjp}. As we are viewing the $(1+1)$-dimensional
boundary as a simplified model for the $(3+1)$-dimensional boundary,
we shall seek a single-component interpretation for the boundary
matter.

Consider in particular the situation in which $f$ and $g$ are both
non-zero and have the same sign. We can then write the energy-momentum
tensor in the perfect fluid form $T_\mn=(\epsilon+p)u_\mu
u_\nu+pg_\mn^{(0)}$, where the normalised future-pointing timelike
velocity vector $u^\mu$ reads
\be
u_\mu=\left(-\left({\gp\over 4\fm}\right)^{1/4},
-\left({\fm\over 4\gp}\right)^{1/4}\right)
\label{umures}
\ee
and the energy density $\epsilon$ and the pressure $p$ are given by
\be
\epsilon=p=\pm\frac{\CL}{8\pi G_3}\sqrt{\gp \fm}
\, ,
\label{epsres}
\ee
with the sign in \nr{epsres} being that of $f$ and~$g$.
Note that the dimensions in these formulas
are correct since $\CL/G_3$ is dimensionless and $f$ and $g$
each have dimension $1/{(\textrm{distance})}^2$.
To find the trajectories of the comoving fluid elements, we have to
integrate the equation
$dx^\mu/d\tau=u^\mu(x^+,x^-)$.
Using the normalisation condition
$1 = -u^2=2u^+u^-=2{(dx^+/d\tau)}^2 (dx^-/dx^+)$,
we find $u^\mu= \left(\sqrt{dx^+/2dx^-},\sqrt{dx^-/2dx^+} \, \right)$,
and comparison with
\nr{umures} then shows that the trajectories are obtained by integrating
the separable differential equation
\be
{dx^-\over dx^+}=\sqrt{{\gp\over\fm}} \, .
\label{paths}
\ee

\section{Special cases}
\label{sec:special}

While the general bulk solution \nr{gensollc} involves the two arbitrary functions
$f$ and~$g$, Einstein's equations \nr{einstein-equations} imply that
this solution must be locally isometric to~AdS$_3$. The solution
can therefore be brought at least locally to the standard Ba\~nados-Teitelboim-Zanelli
(BTZ) form~\cite{henneaux,carlip},
\be
ds^2 =
- F dt^2
+ \frac{dr^2}{F}
+ r^2 {(d\varphi + N^\varphi dt)}^2 ,
\label{gen-BTZ}
\ee
where
\be
F =
\frac{(r^2 - r_+^2) (r^2 - r_-^2)}{\CL^2 r^2} ,
\qquad
N^\varphi =
- \frac{r_+ r_-}{\CL r^2} ,
\ee
and the
the parameters $r_\pm$ determine the mass parameter $M$ and the
angular momentum parameter $J$ by
\be
M =
\frac{r_+^2 + r_-^2}{\CL^2} ,
\qquad
J =
\frac{2 r_+ r_-}{\CL} .
\ee
We shall now write out this transformation for certain
choices for $f$ and~$g$.
These choices will emerge in section
\ref{sec:collision} as regions of special interest in a solution that describes a heavy
ion collision process on the boundary.

\subsection{Bjorken similarity flow}
\label{subsec:bjorken}

Consider the region in which $x^\pm>0$, and let
\be
g(x)=f(x)={M-1\over 4 x^2} \, ,
\label{gfusual}
\ee
where $M$ is a constant.
For $M\ne1$, we have from~\nr{umures}
\ba
u_\mu
&=&
\left(-\sqrt{{x^-\over2 x^+}},-\sqrt{{x^+\over2 x^-}}\right)
={1\over\sqrt2}(-e^{-\eta},
-e^\eta) ,
\quad
u^\mu = {x^\mu\over\tau}={1\over\sqrt2}(e^\eta,e^{-\eta}),
\label{simflow}
\ea
and
\nr{epsres} gives
\be
\epsilon=\frac{\CL}{8\pi G_3}{M-1\over 4x^+x^-}
=\frac{\CL}{16\pi G_3}{M-1\over \tau^2} .
\label{simileps}
\ee
In the coordinates $(t,x)$, \nr{simflow} reads
$u^\mu =(t/\tau,x/\tau)$.
The boundary flow is thus Bjorken similarity flow~\cite{Bjorken:1982qr}.

In Milne coordinates $(\tau,\eta,z)$, the bulk metric reads
\be
ds^2=\frac{\CL^2}{z^2}
\left[-\left(1-\frac{(M-1)z^2}{4\tau^2}\right)^2
d\tau^2+ \left(1+\frac{(M-1)z^2}{4\tau^2}\right)^2
\tau^2 d\eta^2+dz^2\right] .
\label{t-solution}
\ee
This bulk metric was analysed from the
gauge/gravity duality viewpoint in~\cite{klt}. The metric can be transformed to the
BTZ form \nr{gen-BTZ} with $J=0$,
\be
ds^2 =- \left( \frac{r^2}{\CL^2} - M \right) dt^2
+ \frac{dr^2}{ r^2/\CL^2 - M }
+ r^2 d\eta^2 .
\label{btz}
\ee
For positive~$M$, the Killing vector $\partial_t$
has a Killing horizon at $r = \CL \sqrt{M}$,
and the temperature associated with this horizon is $T=\sqrt{M}/(2\pi\CL)$
and the Bekenstein-Hawking entropy
$\Delta S_{\mathrm{BTZ}}$ for an interval $\Delta \eta$ is given by
$
\Delta S_{\mathrm{BTZ}} / (\CL \Delta \eta) =
\sqrt{M} / (4 G_3)$.
A~time-dependent scaling argument from the boundary of
\nr{btz} to the boundary of \nr{t-solution} then gives the flow
\nr{simflow}--\nr{simileps} the time-dependent temperature and entropy density
\be
T(\tau) =\frac{\sqrt{M}}{2\pi\tau} ,
\qquad
s(\tau)
=\frac{\sqrt{M}}{4 G_3}{\CL\over\tau} .
\label{Ttau}
\ee
The term
$-1$ in the factor $M-1$ in \nr{simileps} can be interpreted as a vacuum energy contribution,
and once this contribution is subtracted, the fluid is a perfect fluid in adiabatic expansion
for any positive value of~$M$.

\subsection{Angular momentum}

Let $f(x)$ and $g(x)$ be again proportional to $1/x^2$, but with coefficients that are
not necessarily equal. We write the coefficients as
\be
\gp={M-1-J/\CL\over 4{(x^+)}^2},\qquad \fm={M-1+J/\CL\over 4{(x^-)}^2},
\ee
where $M$ and $J$ are constants. In the domain $x^\pm>0$, going to Milne coordinates
$(\tau,\eta,z)$ then puts the metric in the form investigated in section 5 of \cite{klt},
and it follows from the analysis therein that the metric can be brought to the
spinning BTZ form~\nr{gen-BTZ}.
If $J/\CL = M-1$, so that $g$ vanishes and we have on the boundary a null dust
\nr{dust} moving in the $x^+$ direction, this conclusion can be extended from the
domain $x^\pm>0$ to the half-space $x^- >0$.
Similarly, if $- J/\CL = M-1$, so that $f$ vanishes and we have on the boundary a
null dust \nr{dust} moving in the $x^-$ direction, this conclusion holds in the
half-space $x^+ >0$.

\subsection{Plateau: constant $f$ and $g$}
\label{subsec:fg-constants}

Let now both $f$ and $g$ be constants.

Suppose first that $f$ and $g$ either have the same sign or are both zero. By a boost
in $(x^+,x^-)$ it is then always possible to make $f$ and $g$ equal, so it suffices to
consider this case. We set $f = g = \mu/(2\CL^2)$, where $\mu$ is a dimensionless constant.
The bulk metric reads
\be
ds^2 = \frac{\CL^2}{z^2}
\left\{
- 2 \left(
1 +
\frac{\mu^2 z^4}{16 \CL^2}
\right)dx^+ \, dx^- +
\frac{\mu z^2}{2\CL^2}
\left[
{(dx^+)}^2 +  {(dx^-)}^2 \right]
+dz^2
\right\} ,
\label{f=g-metric}
\ee
and transforming to the coordinates
$(t,x,z)$, brings this metric to the form
\be
ds^2 =
- \left( \frac{\CL}{z} - \frac{\mu z}{4\CL} \right)^2 dt^2
+ \CL^2 \frac{dz^2}{z^2}
+ \left( \frac{\CL}{z} + \frac{\mu z}{4\CL} \right)^2 dx^2 .
\ee
Writing finally
\be
r = \CL \left( \frac{\CL}{z} + \frac{\mu z}{4\CL} \right)  ,
\ee
the metric is transformed to the spinless BTZ form \nr{btz} with
$d\eta\to dx/\CL$ and
$M=\mu$.
If $\mu>0$, the
temperature
associated with the Killing vector $\partial_t$ is
\be
T=\frac{\sqrt{\mu}}{2\pi\CL} ,
\label{T-plateau}
\ee
and the entropy $\Delta S_{\mathrm{BTZ}}$
for an interval $\Delta x$ is
\be
{\Delta S_{\mathrm{BTZ}}\over \Delta x}=
\frac{\sqrt{\mu}}{4 G_3} .
\label{s-BTZ}
\ee

If $f$ and $g$ have opposite sign, a similar analysis can be given, but this case will
not occur in section \ref{sec:collision} below.

Suppose then that $g=0$ but $f\ne0$, so that on the boundary we have the null dust
\nr{dust} moving in the $x^+$ direction.
We now write $f = \mu/\CL^2$, where $\mu$ is a nonvanishing constant. The bulk metric is
\ba
ds^2&=& \frac{\CL^2}{z^2}
\left[ - 2dx^+ \, dx^-+
\frac{\mu z^2}{\CL^2}
{(dx^-)}^2+dz^2  \right]
\nonumber
\\
&=& - \frac{(\CL^4/z^4)}{\CL^2/z^2 + \mu/2} dt^2
+ \CL^2 \frac{dz^2}{z^2}
+ \left( \frac{\CL^2}{z^2} + \frac{\mu}{2} \right)
\left(dx - \frac{(\mu/2)dt}{\CL^2/z^2 + \mu/2} \right)^2
\ea
and can, by introducing the coordinates $(t,r,x)$ with
\be
r = \CL \sqrt{\frac{\CL^2}{z^2} + \frac{\mu}{2}} \, ,
\ee
be written in the form
\be
ds^2 =
- \frac{\left( r^2 - \mu \CL^2/2 \right)^2}{\CL^2 r^2} dt^2
+ \frac{\CL^2 r^2 dr^2}{\left( r^2 - \mu \CL^2/2 \right)^2}
+ r^2 \left( \frac{dx}{\CL} - \frac{(\CL\mu/2) dt}{r^2} \right)^2 .
\ee
This metric is of the BTZ form \nr{gen-BTZ} with $d\varphi \to dx/\CL$,
$M = \mu$ and $J = \mu \CL$. For $\mu>0$ the metric is hence an extremal rotating
BTZ hole, with the angular dimension unwrapped.

A similar analysis applies to the case in which
$f=0$ but $g\ne0$, with $M$ and $J$ now having opposite signs.

\subsection{Interpolation}
\label{subsec:interpolation}

Finally, we set
\be
f(x^-) = \frac{1}{ v^2 a^2 } , \qquad
g(x^+) = \frac{1}{v^2 {(x^+)}^2} ,
\ee
where $a$ is a positive constant of dimension length,
$v$~is a positive dimensionless constant and we assume $x^+>0$.
This case can be regarded as being half-way
between the Bjorken similarity flow case \nr{gfusual}
and the plateau case of subsection~\ref{subsec:fg-constants}.
Note that $\partial_{x^-}$ is a spacelike Killing vector in the bulk,
and the boundary flow \nr{umures} is invariant
under the corresponding null Killing vector $\partial_{x^-}$ on the boundary.

We first transform from $(x^+, x^-, z)$ to $(w^+, w^-, \alpha)$ by
\ba
x^- &=&
v a  w^- ,
\nonumber
\\
\log \! \left(
\frac{2 a v^2 x^+}{\CL^2} \right)
&=&
\frac{v}{\sqrt{1 + {(v/2)}^2}} \, w^+ ,
\nonumber
\\
\log \! \left(
\frac{z}{\CL} \right)
&=&
-\alpha +
\frac{(v/2)}{\sqrt{1 + {(v/2)}^2}} \, w^+ .
\ea
The conformal boundary is then at $\alpha \to \infty$ with fixed $w^\pm$.
These coordinates are adapted to the commuting bulk Killing vectors
$\partial_{w^-} = a v \, \partial_{x^-}$ and $\partial_{w^+}$.
To eliminate from the metric a cross term proportional to~$d\alpha dw^+$, we transform to
$(y^+, y^-, \rho)$, where
\ba
\cosh 2\alpha
&=&
\rho \sqrt{1 + {(v/2)}^2} ,
\nonumber
\\
dw^+ &=&
dy^+ -
\frac{(v/4) \, d\rho}{(\rho^2-1) \, \sqrt{\left[ 1 + {(v/2)}^2 \right] \rho^2 - 1}} ,
\nonumber
\\
dw^- &=&
dy^- -
\frac{(v/4) \, \rho d\rho}{(\rho^2-1) \, \sqrt{\left[ 1 + {(v/2)}^2 \right] \rho^2 - 1}} ,
\ea
and $\rho>1$. The metric becomes
\be
ds^2 =
\CL^2
\left[
{(dy^+)}^2 + {(dy^-)}^2
- 2 \rho \, dy^+ dy^-
+
\frac{d\rho^2}{4 \left(\rho^2-1\right)}
\right] .
\label{eq:y-rho-metric}
\ee
Note that both $a$ and $v$ have now disappeared.
Transforming further to $(\theta^1, \theta^2, \chi)$ by
\ba
y^\pm
&=&
{\textstyle\frac12} (\theta^1 \pm \theta^2 ) ,
\nonumber
\\
\rho &=&  \cosh 2\chi ,
\ea
where $\chi>0$, we have
\be
ds^2 =
\CL^2
\left[ - {(\sinh \chi)}^2 {(d\theta^1)}^2
+ {(\cosh \chi)}^2 {(d\theta^2)}^2
+ d\chi^2
\right] .
\label{eq:theta-chi-metric}
\ee
The metric \nr{eq:theta-chi-metric} is recognised as part of the
AdS$_3$ hyperboloid in coordinates in which the Killing vectors $\partial_{\theta^1}$ and
$\partial_{\theta^2}$ generate two commuting boosts~\cite{henneaux,carlip}: writing the
hyperboloid in embedding coordinates $(T^1, T^2, X^1, X^2)$ as
\ba
\CL^2 &=& {\left(T^1\right)}^2 + {\left(T^2\right)}^2
- {\left(X^1\right)}^2 - {\left(X^2\right)}^2  ,
\nonumber
\\
ds^2 &=& - {\left(dT^1\right)}^2 - {\left(dT^2\right)}^2
+ {\left(dX^1\right)}^2 + {\left(dX^2\right)}^2  ,
\ea
we can choose the transformation to \nr{eq:theta-chi-metric} so that
$\partial_{\theta^1} = T^1 \partial_{X^1} + X^1 \partial_{T^1}$
and
$\partial_{\theta^2} = T^2 \partial_{X^2} + X^2 \partial_{T^2}$.
As $\partial_{x^-} = {(va)}^{-1} \partial_{y^-}
= {(va)}^{-1} \left( \partial_{\theta^1} - \partial_{\theta^2} \right)$, it finally follows that
\be
\partial_{x^-} = \frac{1}{va}
\left(
T^1 \partial_{X^1} + X^1 \partial_{T^1}
- T^2 \partial_{X^2} - X^2 \partial_{T^2}
\right) .
\label{eq:mix-killing-invariant}
\ee

To summarise, the invariant geometric characterisation of the bulk Killing vector
$\partial_{x^-}$ can be read off from~\nr{eq:mix-killing-invariant}.
As $\partial_{x^-}$ is the bulk continuation of the boundary Killing vector under
which the fluid flow is invariant, one is tempted to seek thermodynamics for the
boundary flow from thermodynamics of bulk phenomena that are invariant under~$\partial_{x^-}$.
This strategy works for Bjorken similarity flow, as outlined in subsection
\ref{subsec:bjorken}~\cite{klt}, and for the plateau flow, as outlined in
subsection~\ref{subsec:fg-constants}, since in both cases the Killing vector
in question is a spacelike boost and the Killing horizon of the commuting
orthogonal boost has a Hawking temperature.
With the Killing vector~\nr{eq:mix-killing-invariant}, however, a thermodynamical
bulk interpretation appears not to be known, and there is evidence that one may
not exist~\cite{Balasubramanian:2003kq}. It is certainly possible to transform
the bulk metric to the BTZ form: given any $M$ and $J$ satisfying $M > |J/\CL|$,
the transformation
\ba
y^\pm
&=&
\frac12
\sqrt{M \mp (J/\CL)}
\left( \frac{t}{\CL} \pm \varphi \right) ,
\nonumber
\\
\rho &=&
\frac{2 {(r/\CL)}^2 - M}{\sqrt{M^2 - {(J/\CL)}^2}} ,
\ea
takes the metric \nr{eq:y-rho-metric} to the BTZ metric~\nr{gen-BTZ}. However, we then have
\be
\partial_{y^\pm} =
\frac{\CL \partial_t \pm \partial_\varphi}{\sqrt{M \mp (J/\CL)}} ,
\label{eq:party-partBTZ}
\ee
and consideration of the Killing vector
$\partial_{x^-} = {(va)}^{-1} \partial_{y^-}$
does not appear to single out specific values of $M$ and $J$
for which the Killing horizon of $\partial_t$ could be argued
to provide a temperature for the boundary flow.

\section{Collision}
\label{sec:collision}

We now turn to situations in which a collision of two null dust flows
evolves into Bjorken similarity flow at late times. We hence assume both
$f$ and $g$ to vanish at negative argument, be nonzero and have the same sign for positive
argument and approach Bjorken similarity flow form
\nr{gfusual} at large positive argument.

\begin{figure}[!tb]
\begin{center}

\vspace{-0.8cm}
\includegraphics[width=0.6\textwidth]{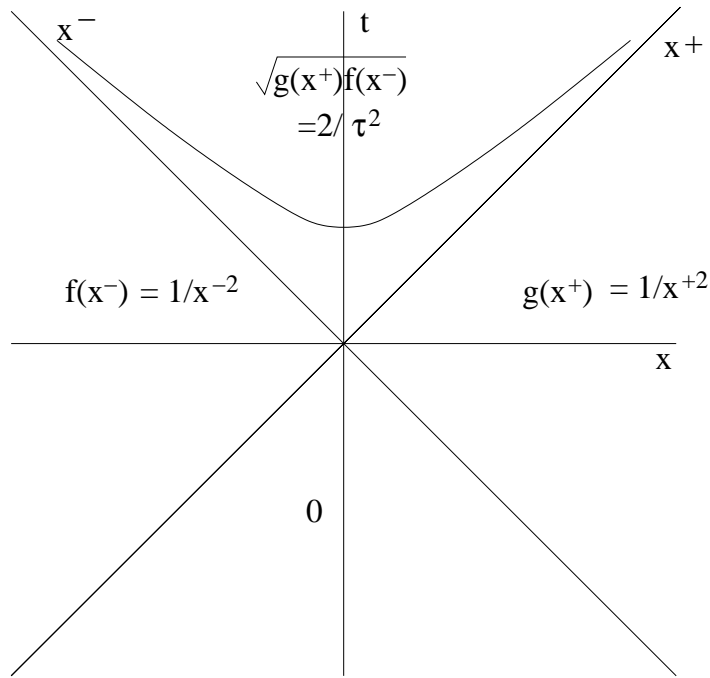}
\end{center}
\caption{\small Structure of the conformally invariant boundary
flow following from the
solution \nr{gensollc}, assuming that
$f$ and $g$ are vanishing at negative argument and are nonvanishing and have the same sign
for positive argument. Within the past light cone of the origin the energy-momentum
tensor vanishes. The left wedge
$x^+ < 0 < x^-$ contains null dust moving to the right, and the right wedge
$x^- < 0 < x^+$ contains null dust moving to the left. The future light cone of
the origin contains expanding matter with the flow given by \nr{umures} and energy density
given by~\nr{epsres}.
The special solution \nr{gfunreg} that gives rise to unregulated Bjorken similarity
flow \nr{simflow}--\nr{simileps} within the future light cone of the origin is also
indicated; for this solution the energy density \nr{simileps} diverges as $\tau\to0$.}
\label{bjflow}
\end{figure}

The general structure of the boundary flow is shown in Figure~\ref{bjflow}.
Within the past light cone of the origin we have vacuum, the left wedge
$x^+ < 0 < x^-$ contains null dust moving to the right, and the right wedge $x^- < 0 < x^+$
contains null dust moving to the left. Within the future light cone of the origin the null
dusts have collided and formed the flow given by
\nr{umures} and~\nr{epsres}, which at late times tends to
Bjorken similarity flow,
\nr{simflow} and~\nr{simileps}.

The key point is now that a suitable choice of the null dust profiles resolves the initial
singularity of Bjorken similarity flow. We look at unregulated Bjorken similarity flow in
subsection
\ref{subsec:unregulated} analyse two regulated versions in subsections
\ref{subsec:smooth} and~\ref{subsec:plateau}.

\subsection{Unregulated Bjorken similarity flow}
\label{subsec:unregulated}

Consider first the choice
\be
g(x)=f(x)={M-1\over 4 x^2} \, \Theta(x) ,
\label{gfunreg}
\ee
where $M$ is a constant. Within the future light cone of the origin we then have
Bjorken similarity flow \nr{simflow}--\nr{simileps}. As reviewed in
subsection~\ref{subsec:bjorken}, for positive $M$ this flow can be understood to
describe a perfect fluid in adiabatic expansion with $\epsilon(\tau)\sim 1/\tau^2$
and $T(\tau)\sim 1/\tau$.

The problem with this
scenario is however that the energy density diverges at $\tau\to0$, and
the energies $E^\pm$ \nr{etotplusminus} of the initial null dust components also diverge.
While for a (3+1)-dimensional boundary the bulk dynamics may provide criteria
for regulating the initial singularity~\cite{kovchegov,Benincasa:2007tp},
we are not aware of similar criteria for (1+1)-dimensional boundary.
We shall therefore regulate the singularity by a bottom-up phenomenological
approach, introducing in the functions $f$ and $g$ and new scale that corresponds
to the initial conditions.

\subsection{Smooth profile}
\label{subsec:smooth}

As a first alternative, we modify Bjorken similarity flow functions
\nr{gfunreg} to
\be
f(x)=g(x)= \frac{M-1}{4 \left( x^2+a^2 \right)} \, \Theta(x),
\label{reg1}
\ee
where $a$ is a positive constant of dimension length. As the only other length
scale in the problem is~$\CL$, one may expect on phenomenological grounds $a$
to be proportional to $\CL$ by a dimensionless constant of order unity.
For $x^\pm>0$, we then obtain from \nr{epsres}
\be
\epsilon=p=\frac{\CL}{8\pi G_3}{\fra14(M-1)\over\sqrt{(x^{+2}+a^2)(x^{-2}+a^2)}}=
\frac{\CL(M-1)}{16\pi G_3}{1\over\sqrt{\tau^4+2a^2\cosh (2\eta)+4a^4}}.
\label{eq:eps-smreg}
\ee
The energy density thus remains finite at $\tau\to0$,
has some $\eta$ dependence at intermediate $\tau$
and approaches the similarity solution \nr{simileps}
as $\tau\to\infty$ at fixed~$\eta$.

\begin{figure}[!tb]
\begin{center}
\vspace{-0.8cm}
\includegraphics[width=0.6\textwidth, angle=45]{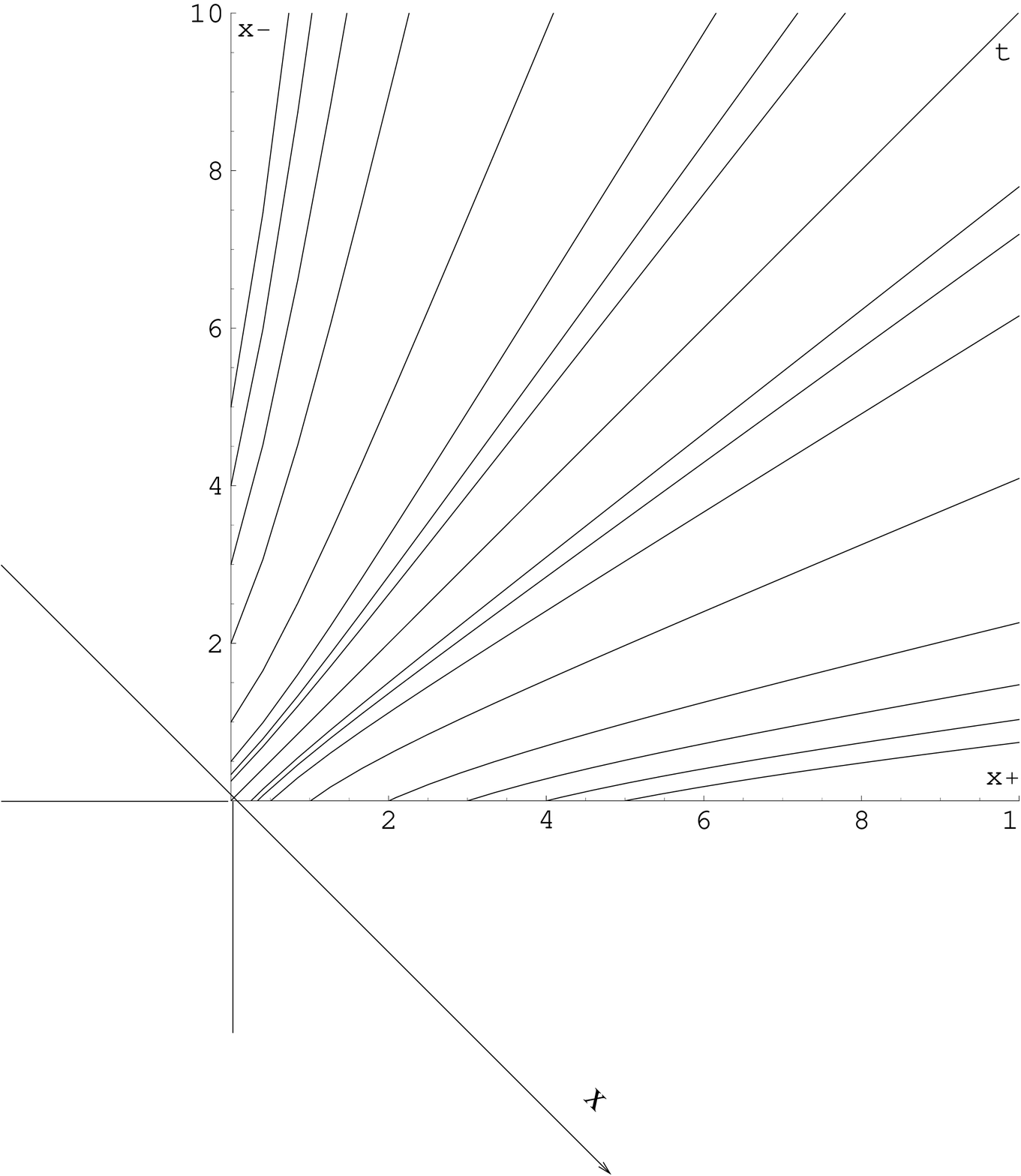}
\end{center}
\vspace{-1.6cm}
\caption{\small
Flow lines \nr{pathsex}
for $a=1$ and
$x_i=1/4$, $1/3$, $1/2$, $1$, $2$, $3$, $4$ and~$5$,
and the corresponding flow lines with $x^+$ and $x^-$ interchanged.}
\label{radat}
\end{figure}

To see
how the singularity at $\tau\to0$ is opened up, we may compute the flow lines.
Now all the particles do not emanate from the origin $x^+=x^-=0$.
When the initial point is on the right-hand-side branch of the light cone, at
$x^-=0$ and $x^+=x_i>0$, equation \nr{paths} implies
\be
\int_0^{x^-}{dx^-\over\sqrt{{(x^{-})}^2+a^2}}=\int_{x_i}^{x^+}
{dx^+\over\sqrt{{(x^{+})}^2+a^2}} ,
\ee
which can be integrated to
\be
x^-(x^+)
=
\frac{{\left(x^++\sqrt{{(x^{+})}^2+a^2} \, \right)}^2
-
{\left(x_i+\sqrt{x_i^2+a^2} \, \right)}^2}
{2 \left(x^++\sqrt{{(x^{+})}^2+a^2} \, \right)
\left(x_i+\sqrt{x_i^2+a^2} \, \right)}.
\label{pathsex}
\ee
Note that these paths stay in the region $x^+ > x^-$.
When the initial point is on the left-hand-side branch of the light cone, the
situation is similar with $x^+$ and $x^-$ interchanged. The only path that starts
from the origin is $x^+ = x^-$, obtained from \nr{pathsex} with $x_i=0$.
A~selection of flow lines is shown in Figure~\ref{radat}.
In the limit $a\to0$, we obtain Bjorken similarity flow~\nr{simflow}.

What this regularisation seems to lack, however, is a way to associate a
temperature to the early stages of the flow. With Bjorken similarity flow,
the temperature arose by noting that the flow is invariant under longitudinal
boosts generated by the Killing vector~$\partial_\eta$, the bulk solution
\nr{t-solution} extends this Killing vector from the boundary to the bulk,
and in the bulk metric \nr{btz} there is a commuting timelike Killing vector
$\partial_t$ whose horizon has a Hawking temperature~\cite{klt}.
We have not found a similar isometry
argument that would apply to the flow with the particle paths~\nr{pathsex}.

\subsection{Hard core profile}
\label{subsec:plateau}

As a second alternative, we modify Bjorken similarity flow functions
\nr{gfunreg} to
\be
f(x) = g(x) = {M-1\over4}\left[{1\over x^2}\Theta(x-a)
+{1\over a^2}\Theta(x)\Theta(a-x)\right],
\label{reg2}
\ee
where $a$ is again a positive constant of dimension length.
Phenomenologically, we might call the second term in \nr{reg2}
a hard core ``nucleus'', of characteristic size~$a$,
and the first term its
longitudinal ``tail''. We shall assume $M>1$.

\begin{figure}[!tb]
\begin{center}

\vspace{-0.8cm}
\includegraphics[width=0.6\textwidth]{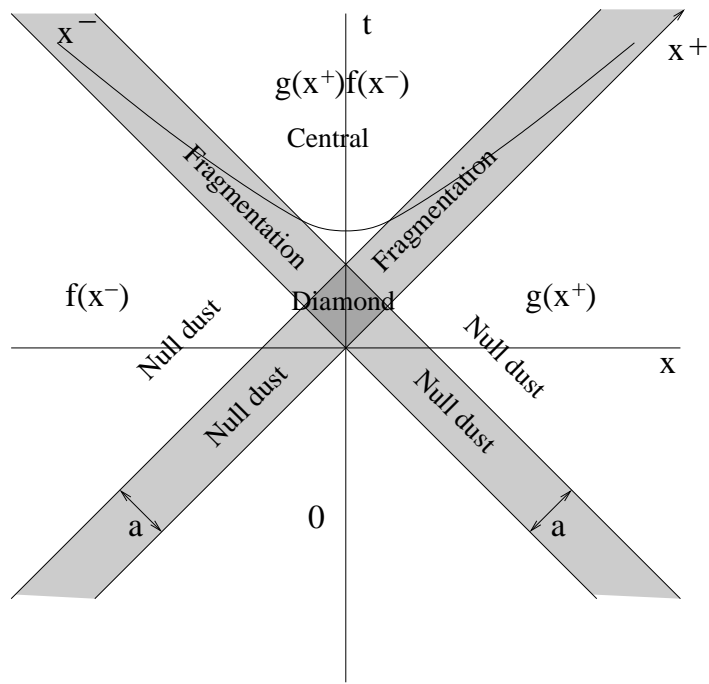}
\end{center}

\vspace{-0.8cm}
\caption{\small The boundary flow with the regularisation~\nr{reg2}.
The various regions are described in the text.
A~curve of constant $\tau$ at $\tau > a\sqrt{2}$ is shown in the region
$x^\pm >0$.
}
\label{figreg2}
\end{figure}

The different regions on the boundary are shown in Figure~\ref{figreg2}.
From \nr{etotplusminus} we find that the
``incident beam energies'' in the left and right wedges are given by
\be
E^\pm =
{\CL(M-1)\over 32\pi G_3}{\sqrt2\over a} \, ,
\label{beamenergy}
\ee
and for each null dust component this energy is distributed equally between the nucleus
and its tail. As shown in subsection~\ref{subsec:fg-constants}, the bulk metric in each
null dust nucleus region can be thought of an extremal BTZ black hole, with the angular
dimension unwrapped,
and has therefore vanishing temperature.
We thus regard the incident null dust beams as completely unthermalised, both in the
nucleus and in the tail.

\begin{figure}[!tb]
\begin{center}

\includegraphics[width=0.45\textwidth]{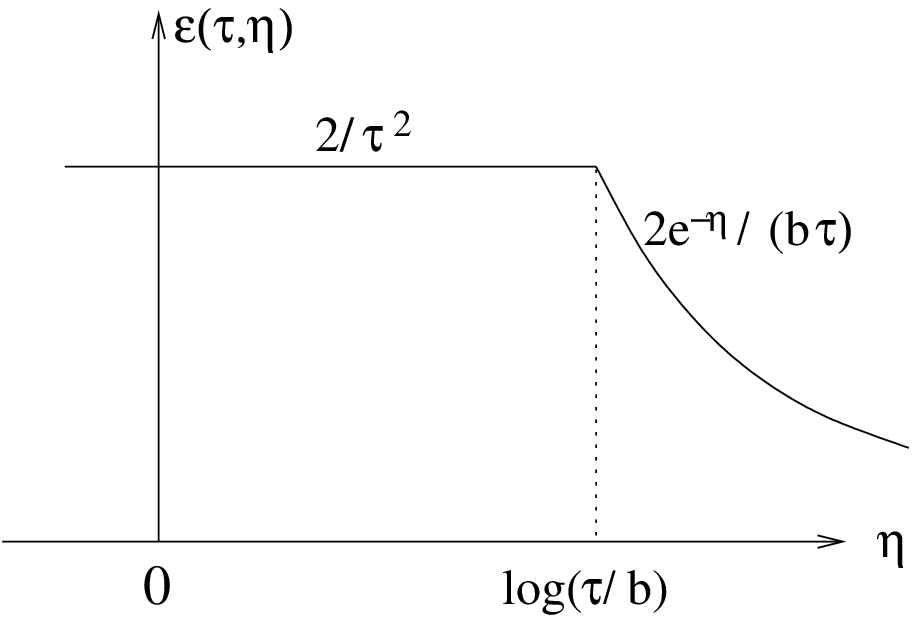}\hspace{0.8cm}
\includegraphics[width=0.45\textwidth]{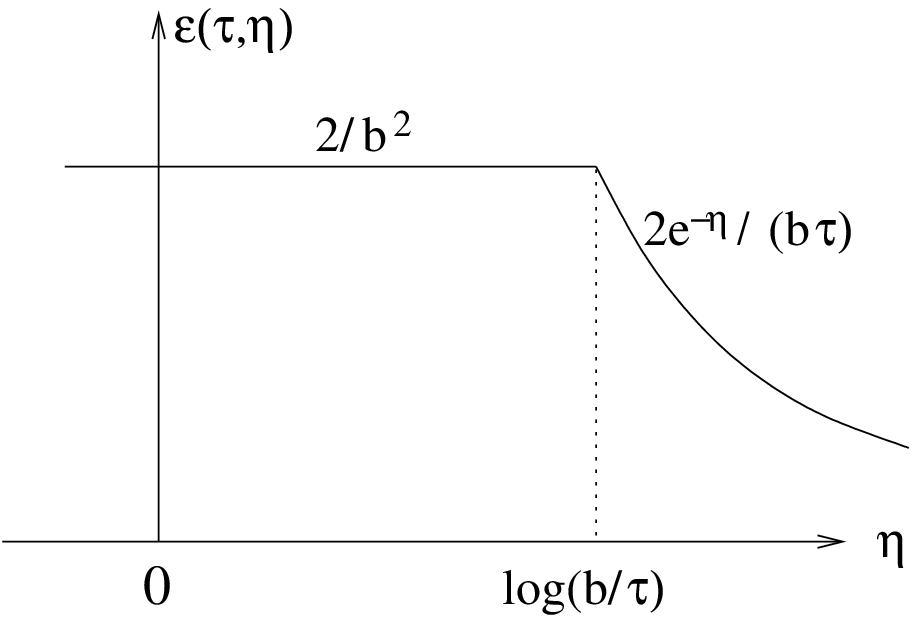}
\end{center}
\caption{\small The energy density \nr{epstaueta1}
divided by $\CL(M-1)/(32\pi G_3)$
as a function of $\eta$ at constant $\tau$ in the
regularisation~\nr{reg2}. The left panel shows the case
$\tau>a\sqrt2$ and the right panel the case $\tau<a\sqrt2$. The parameter $b$ equals $a\sqrt2$.}
\label{figreg3}
\end{figure}

The region $x^\pm>0$, in which the beams have collided, is divided into the central
region in which the null dust tails overlap, two fragmentation regions in which one
nucleus overlaps the other's tail, and the interaction diamond in which the two nuclei overlap.
In the light cone coordinates, the velocity field is given by
\be
u^\mu =
\left\{
\begin{array}{ll}
\displaystyle \left(\sqrt{x^+\over 2x^-},\sqrt{x^-\over 2x^+} \> \right)
= {x^\mu\over\tau} ,
& \textrm{central;}
\\[4ex]
\displaystyle \left(\sqrt{x^+\over 2a},\sqrt{a\over 2x^+} \, \right) ,
& \textrm{fragmentation at $x^- < a$;}
\\[4ex]
\displaystyle \left(\sqrt{a\over 2x^-} , \sqrt{x^-\over 2a} \, \right) ,
& \textrm{fragmentation at $x^+ < a$;}
\\[4ex]
\displaystyle \left({1\over\sqrt2},{1\over\sqrt2}\right) ,
& \mathrm{diamond.}
\end{array}
\right.
\label{velfield}
\ee
Note that in the interaction diamond the fluid is thus
at rest in the Minkowski coordinates $(t,x)$.
For the energy density,
\nr{epsres} and \nr{reg2} give
\ba
\epsilon(x^+,x^-)&=&{\CL(M-1)\over 32\pi G_3}
\left[{1\over x^+x^-}\Theta(x^+-a)\Theta(x^--a) \right.
\nonumber
\\[1ex]
&&
\qquad\qquad\quad
\left.+{1\over ax^+}\Theta(x^+-a)\Theta(x^-)\Theta(a-x^-)+
(x^+\leftrightarrow x^-)\right.
\nonumber
\\[1ex]
&&
\qquad\qquad\quad
\left.+{1\over a^2}\Theta(a-x^+)\Theta(x^+)\Theta(x^-)\Theta(a-x^-)\right] ,
\label{epspm}
\ea
or, in Milne coordinates,
\ba
\epsilon(\tau,\eta)&=&{\CL(M-1)\over 32\pi G_3}
\left\{
\Theta \! \left(\tau-a\sqrt2 \, \right)
\left[
{2\over\tau^2}
\Theta \! \left( \log{\tau\over a\sqrt2}-|\eta| \right)
+ {\sqrt{2} \, e^{-|\eta|} \over a\tau}
\Theta \! \left( |\eta| - \log{\tau\over a\sqrt2}\right)
\right]
\right.
\nonumber
\\[1ex]
&&
\qquad\quad
\left.+
\Theta \! \left(a\sqrt2-\tau\right)
\left[
{1\over a^2}
\Theta \! \left(\log{a\sqrt2\over\tau} - |\eta| \right)
+
{\sqrt{2} \, e^{-|\eta|} \over a\tau}
\Theta \! \left( |\eta| - \log{a\sqrt2 \over\tau} \, \right)
\right]
\right\}.
\nonumber
\\[1ex]
&&
\label{epstaueta1}
\ea
In \eq\nr{epspm} the first line describes the central region,
the second line the two fragmentation
regions and the last line the interaction diamond.
In~\eq\nr{epstaueta1}, plotted in
Figure~\ref{figreg3},
the first line gives
the energy density for
$\tau>a\sqrt2$, where the interaction diamond does not contribute, and the second
line gives the energy density for
$\tau<a\sqrt2$, where the central region does not contribute.
At $\eta=0$, \nr{epstaueta1} reduces to
\be
\epsilon(\tau,\eta=0)={\CL(M-1)\over 32\pi G_3}
\left[
\frac{2 \, \Theta \! \left(\tau-a\sqrt2 \, \right)}
{\tau^2}
+ \frac{\Theta \! \left(a\sqrt2-\tau\right)}
{a^2}
\right] .
\label{epstaueta=0}
\ee

\begin{figure}[!tb]
\begin{center}

\includegraphics[width=0.7\textwidth]{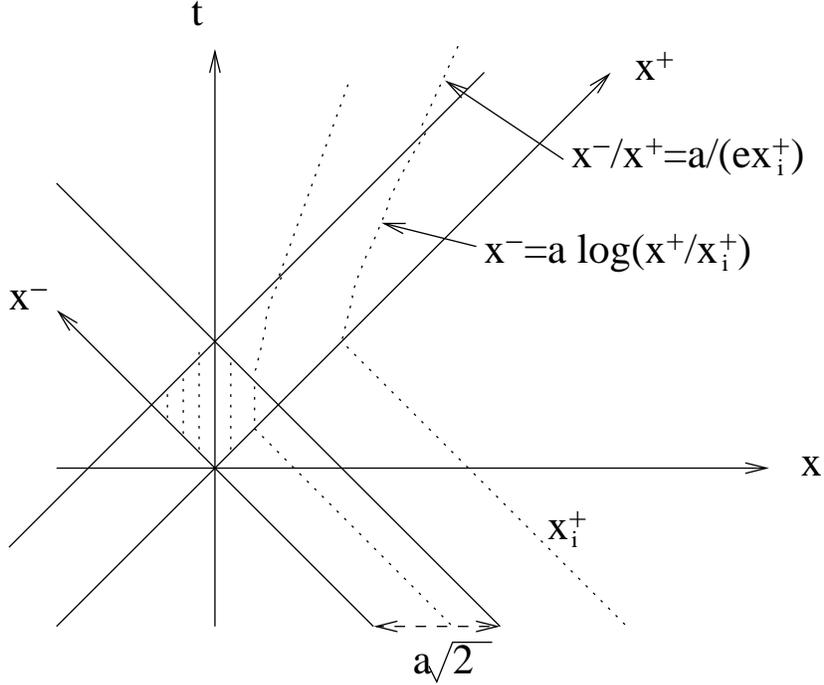}
\end{center}
\caption{\small Paths
\nr{xilessa} and \nr{ximorea}
of the fluid elements moving initially to the left with $x^+=x^+_i$.
In the Minkowski coordinates $(t,x)$, the fluid comes
momentarily to rest on the light cone of the origin,
and it stays at rest everywhere within the interaction diamond.}
\label{figpartpaths}
\end{figure}

We may again obtain a concrete view of how the singularity is opened up by
computing the flow lines.
For concreteness, consider the flow lines approaching from the right. In
the null dust region these lines
are the null lines $x^+ = x^+_i$, where $x^+_i$ is a positive constant.
The lines continue into the other regions for $x^+_i<a$ as
\be
x^- =
\left\{
\begin{array}{ll}
\displaystyle x^+-x^+_i , \ \textrm{or} \ x = x^+_i /\sqrt2 ,
& \textrm{diamond;}
\\
\displaystyle a-x^+_i+a\log{x^+\over a} ,
& \textrm{fragmentation region;}
\\
\displaystyle x^+ e^{x^+_i/a} ,
& \textrm{central region,}
\end{array}
\right.
\label{xilessa}
\ee
and for $x^+_i>a$ as
\be
x^- =
\left\{
\begin{array}{ll}
\displaystyle a\log{x^+\over x^+_i} ,
& \textrm{fragmentation region;}
\\[3ex]
\displaystyle x^+\, {a\over ex^+_i} ,
& \textrm{central region.}
\end{array}
\right.
\label{ximorea}
\ee
The lines are plotted in Figure~\ref{figpartpaths}.
What is most striking here is that
the fluid is brought momentarily to rest
on the light cone of the origin, and in the fragmentation region it
then accelerates in the direction opposite to
where it came from. The conformal matter is thus extremely opaque. This seems
superficially to be in complete disagreement with what one knows about
heavy ion collisions, where nuclei are extremely transparent. However,
this transparency is due to the valence quarks which represent non-conformal
degrees of freedom in nuclei. By contrast, the conformal degrees of freedom are
those built in the classical gluon fields (see Section~\ref{glasma})
computed from the valence currents as sources~\cite{lappi}.

For further insight into the dynamics, we may
consider the conserved energy \nr{etotplusminus} on a surface of constant~$t$.
For $t<0$, all of the energy is in the incident null dust flows, while at
$t=a/\sqrt2$ exactly half of the energy is in the interaction diamond and
the other half is still in the null dust flows. As $t$ increases, more and
more of the energy moves into the central region: for
$t > a \sqrt2$, the energy in the central region reads
\be
E_{\mathrm{central}}(t) =
\frac{\CL(M-1)}{32\pi G_3 }{\sqrt2\over a}
\left(1-\frac{1}{t\sqrt2/a -1}\right).
\label{eq:ecentralt}
\ee
Note that as $T_{\mu\nu}$ is conserved,
$E_{\mathrm{central}}(t)$ can be alternatively calculated as the
integral of $\cosh \eta$ times the comoving energy density
$\epsilon(\tau,\eta)$
\nr{epstaueta1} over the constant $\tau$ surface that meets the constant $t$ surface
on the boundary of the central region,
$t\sqrt2/a -1 = \tau^2 / (2a^2)$:
\be
E_\rmi{central}(\tau)=2\int_0^{\log(\tau/a\sqrt2)}
d\eta \cosh\eta\,\,\tau\epsilon(\tau>a\sqrt2,\eta)
=
{\CL(M-1)\over 32\pi G_3}{\sqrt2\over a}\left(1-{2a^2\over\tau^2}\right).
\label{epscentral}
\ee
The factor $\cosh\eta$ here comes from the relative velocity of the flow lines,
orthogonal to the constant $\tau$ surfaces, and the vector~$\partial_t$, with
respect to which the conserved energy was defined.

Now, to what extent can we recover time-dependent thermodynamics for the fluid
flow after the beams have collided? The
energy density is given by~\nr{epspm}, or equivalently~\nr{epstaueta1},
and the equation of state is $\epsilon = p$, as is expected of a
(1+1)-dimensional situation. The crucial step to thermodynamics is to
find a reasonable notion of temperature.
In the fragmentation regions the situation is unclear,
for reasons discussed in subsection~\ref{subsec:interpolation}.
In the central region, however, the
flow is Bjorken similarity flow. If boundary effects can be neglected,
the arguments of subsection \ref{subsec:bjorken} give this flow the
$\tau$-dependent temperature and entropy density
\be
T_{\mathrm{central}}=\frac{\sqrt{M}}{2\pi\tau} ,
\qquad
s_{\mathrm{central}}
=\frac{\sqrt{M}}{4 G_3}{\CL\over\tau} .
\label{eq:Tcentral}
\ee
Neglecting the boundary effects is likely to be justified provided
the proper length $\tau \Delta\eta$
of a constant $\tau$ curve through the central region is much larger than the
length scale $T_{\mathrm{central}}^{-1}$ associated with the temperature,
and for values of $|\eta|$ that are not close
to the boundary of the central region. These conditions read
\be
\frac{\pi}{\sqrt{M}}
\ll
\log \frac{\tau}{a\sqrt2} ,
\qquad
|\eta| \ll
\log \frac{\tau}{a\sqrt2} ,
\ee
or equivalently
\be
a \, e^{\pi/\sqrt{M}} \ll \sqrt{x^+ x^-} ,
\qquad
a \ll x^\pm .
\label{eq:val-bjorken}
\ee
Similarly, if the boundary effects in the interaction diamond are neglected,
the arguments of
subsection \ref{subsec:fg-constants} give the static flow therein the constant temperature
and entropy density
\be
T_{\mathrm{diamond}}=\frac{\sqrt{M-1}}{2\sqrt{2} \, \pi a} ,
\qquad
s_{\mathrm{diamond}} =
\frac{\sqrt{M-1}}{4 G_3}
\frac{\CL}{a \sqrt2} ,
\label{eq:Tdiamond}
\ee
and neglecting the boundary effects is likely to be justified provided the diamond
is much larger than the length scale $T_{\mathrm{diamond}}^{-1}$ associated with the temperature
and we are not close to the boundaries of the diamond,
which conditions read
\be
1 \ll M ,
\qquad
\bigl|
x^\pm - {\textstyle\frac12}a
\bigr| \ll a .
\label{eq:val-plateau}
\ee
The formulas \nr{eq:Tcentral} and \nr{eq:Tdiamond}
for the temperature and the entropy density
are thus likely to be reliable whenever $M\gg1$,
which can be regarded as the condition
that the bulk is semiclassical~\cite{malda-stro},
and we are not close to the
boundary of respectively the central region and the interaction diamond.
Note that these conditions do
not depend on how $a$ compares with the other length scales of the problem,
$\CL$ and~$\CL\sqrt{M}$.
As a consistency check,
we note that in the limit of large $M$ we have $\sqrt{M-1} \approx \sqrt{M}$,
and formulas \nr{eq:Tcentral} and \nr{eq:Tdiamond} for
the temperature and the entropy density
then agree at the point $(\tau,\eta)= (a\sqrt2, 0)$, at which the central
region and the interaction diamond meet.

The model provides thus a consistent thermodynamical picture both in the
hard core interaction region immediately after the collision, where the
temperature and entropy density are approximately constant,
and at late times, where the temperature falls off as $1/\tau$ and the
fluid is a perfect fluid in adiabatic expansion~\cite{klt}. In Section
\ref{glasma} we shall compare this (1+1)-dimensional thermodynamics with
what would be expected in 3+1 dimensions in the classical gluon field model with
currents of nuclei as sources.

\section{Ensemble of classical gluon fields}
\label{glasma}

Contrast now the above with the modeling of heavy ion collisions using ensembles of
classical color fields \cite{lappi}. One solves the color fields $A_\mu^a(t,\bfx)$
in suitably chosen gauges from the Yang-Mills equations
\be
[D_\mu,F^{\mu\nu}]=J^\nu ,
\label{eom}
\ee
where the source current represents the two nuclei moving along the light cone,
\be
J^\mu=\delta^{\mu+}\delta(x^-)\rho_{(1)}(\xt)+
\delta^{\mu-}\delta(x^+)\rho_{(2)}(\xt).
\label{current}
\ee
This current exists for all $x^\pm$ and has the nuclear transparency built in it.
The two color charge densities $\rho_{(m)}(\xt)$ are, independently for
the two nuclei, drawn from a Gaussian (in the simplest version of the model) random
ensemble,
\be
\langle \rho^a_{(m)}(\xt)\rho^b_{(m)}(\yt)\rangle=
g^2\mu_{(m)}^2\delta^{ab}\delta^2(\xt-\yt),
\quad m=1,2,
\label{rhorho}
\ee
where $g^2\mu$ is a parameter describing the transverse density of color charges
related \cite{lappigsqmu} to the saturation scale by
$Q_s^2\approx0.36(g^2\mu)^2\approx 0.05{\rm GeV}^2A^{1/3}(\sqrt{s}/Q_s)^{0.3}$.
All information on the size of the nucleus and collision energy is built in $Q_s$.
The computed expectation values $\langle A_\mu^aA_\nu^b\rangle$ can be converted
to multiplicities and initial energy densities.

Numerical results are reported, for example, in \cite{lappiendens}. One finds the
initial energy density $\epsilon(\tau,\eta=0)\sim Q_s^4$ for $\tau\lsim 1/Q_s$; thermalisation
takes place at later times. In view of the different dimensionalities the comparison is
only qualitative, but with the identification $a\sim 1/Q_s$ the behavior is similar
to that in \eq\nr{epstaueta=0}. We thus suggest this interpretation for the parameter $a$
introduced above. The gravity/CFT duality picture has also thermalisation built in.

\section{Conclusions}

We have in this paper presented an exact AdS$_3$ gravity solution
which represents in the bulk
a collision of two plane waves and on
the conformal boundary a collision of two (1+1)-dimensional extended objects.
We chose the incoming wave profiles to consist of a hard core ``nucleus'',
of boundary length scale $a$ that can (but need not) be
comparable to the AdS length scale $\CL$ in the bulk,
and a ``tail'' with a $1/x^2$ falloff.
In the region after the collision we then identified
an interaction diamond where the hard core nuclei overlap,
a central region where the tails overlap,
and two interpolating fragmentation regions.

We saw that the boundary matter \emph{recoils\/} in the collision:
the conformal matter is thus extremely opaque. The interaction
diamond corresponds, qualitatively, to the part of collision dynamics described by
classical Yang-Mills field models with the indentification $a\sim 1/Q_s$,
where $Q_s$ is the saturation scale. There is also some qualitative similarity to
Landau's hydrodynamic model \cite{bjp,landau}.

We identified the bulk duals of the central region and the
interaction diamond as parts of a
spinless BTZ black hole spacetime, with the angular dimension unwrapped,
and in each case the local invariance of the boundary flow under a spacelike
Killing vector was used to identify a horizon and its Hawking temperature in
the bulk. When the colliding matter is sufficiently energetic,
these bulk black holes are semiclassical ($M\gg1$),
and we used the bulk-boundary correspondence to induce from the bulk temperature
a temperature to the boundary flow.
We argued that in the limit $M\gg1$ the transition effects between the regions
are negligible provided one is not close to the boundary of the region.
In this approximation
the flow in the
interaction diamond the is static with constant
temperature and entropy density, and
the flow in the central region is
Bjorken similarity flow of a perfect fluid in adiabatic expansion, with
temperature inversely proportional to the proper time~\cite{klt}.

There are clearly many open issues to be studied.
Within our colliding plane wave solution in the AdS$_3$ bulk,
one question is how to introduce thermodynamics even
when the boundary flow is not invariant under a spacelike Killing vector.
The answer would open an avenue for investigating various
phenomenologically-motivated thermalisation scenarios by suitable choices
of the incoming wave profiles.
The main issue of interest would clearly be to extend
the work to more spatial dimensions, where
gravitational bulk
dynamics can be expected to give more structure to the boundary theory.

\vspace{0.5cm}
Acknowledgements. We thank Esko Keski-Vakkuri and Tuomas Lappi for
discussions, Romuald Janik for correspondence and Kostas Skenderis for
bringing \cite{ss} to our attention. This research has been supported
in part by Academy of Finland, contract number 109720 and by STFC (UK)
grant PP/D507358/1.  JL~thanks Helsinki Institute of Physics for
hospitality at an early stage of the work.


\begin{thebibliography}{99}

\bibitem{jp}
R.~A.~Janik and R.~Peschanski,
  ``Asymptotic perfect fluid dynamics as a consequence of AdS/CFT,''
  Phys.\ Rev.\ D {\bf 73}, 045013 (2006)
  [arXiv:hep-th/0512162].

\bibitem{jp2}
  R.~A.~Janik and R.~Peschanski,
  ``Gauge / gravity duality and thermalization of a boost-invariant perfect
  fluid,''
  Phys.\ Rev.\ D {\bf 74}, 046007 (2006)
  [arXiv:hep-th/0606149].

\bibitem{nakamura1}
  S.~Nakamura and S.~J.~Sin,
  ``A holographic dual of hydrodynamics,''
  JHEP {\bf 0609}, 020 (2006)
  [arXiv:hep-th/0607123].

\bibitem{nakamura2}
S.~J.~Sin, S.~Nakamura and S.~P.~Kim,
  ``Elliptic flow, Kasner universe and holographic dual of RHIC fireball,''
  JHEP {\bf 0612}, 075 (2006)
  [arXiv:hep-th/0610113].

\bibitem{janik1}
R.~A.~Janik,
  ``Viscous plasma evolution from gravity using AdS/CFT,''
  Phys.\ Rev.\ Lett.\  {\bf 98}, 022302 (2007)
  [arXiv:hep-th/0610144].

\bibitem{janik2}
M.~P.~Heller and R.~A.~Janik,
  ``Viscous hydrodynamics relaxation time from AdS/CFT,''
  Phys.\ Rev.\  D {\bf 76}, 025027 (2007)
  [arXiv:hep-th/0703243].

\bibitem{kt}
K.~Kajantie and T.~Tahkokallio,
  ``Spherically expanding matter in AdS/CFT,''
  Phys.\ Rev.\  D {\bf 75}, 066003 (2007)
  [arXiv:hep-th/0612226].

\bibitem{kovchegov}
Y.~V.~Kovchegov and A.~Taliotis,
  ``Early time dynamics in heavy ion collisions from AdS/CFT correspondence,''
  Phys.\ Rev.\  C {\bf 76}, 014905 (2007)
  [arXiv:0705.1234 [hep-ph]].

\bibitem{klt}
  K.~Kajantie, J.~Louko and T.~Tahkokallio,
  ``Gravity dual of 1+1 dimensional Bjorken expansion,''
  Phys.\ Rev.\  D {\bf 76}, 106006 (2007)
  [arXiv:0705.1791 [hep-th]].

\bibitem{Lin:2006rf}
  S.~Lin and E.~Shuryak,
  ``Toward the AdS/CFT gravity dual for High Energy Collisions: I.Falling into
  the AdS,''
  arXiv:hep-ph/0610168v2.

\bibitem{Nakamura:2007nx}
  S.~Nakamura, Y.~Seo, S.~J.~Sin and K.~P.~Yogendran,
  ``Baryon-charge Chemical Potential in AdS/CFT,''
  arXiv:0708.2818 [hep-th].

\bibitem{Hatta:2007cs}
  Y.~Hatta, E.~Iancu and A.~H.~Mueller,
  ``Deep inelastic scattering off a N=4 SYM plasma at strong coupling,''
  arXiv:0710.5297 [hep-th].

\bibitem{Benincasa:2007tp}
  P.~Benincasa, A.~Buchel, M.~P.~Heller and R.~A.~Janik,
  ``On the supergravity description of boost invariant conformal plasma at
  strong coupling,''
  arXiv:0712.2025 [hep-th].

\bibitem{Alsup:2007bs}
  J.~Alsup and G.~Siopsis,
  ``Bjorken flow from an AdS Schwarzschild black hole,''
  arXiv:0712.2164 [hep-th].

\bibitem{Bhattacharyya:2007jc}
  S.~Bhattacharyya, V.~E.~Hubeny, S.~Minwalla and M.~Rangamani,
  ``Nonlinear Fluid Dynamics from Gravity,''
  arXiv:0712.2456 [hep-th].

\bibitem{ssz}
  E.~Shuryak, S.~J.~Sin and I.~Zahed,
  ``A gravity dual of RHIC collisions,''
  J.\ Korean Phys.\ Soc.\  {\bf 50}, 384 (2007)
  [arXiv:hep-th/0511199].

\bibitem{matschull}
  H.~J.~Matschull,
  ``Black hole creation in 2+1-dimensions,''
  Class.\ Quant.\ Grav.\  {\bf 16}, 1069 (1999)
  [arXiv:gr-qc/9809087].

\bibitem{holst}
  S.~Holst and H.~J.~Matschull,
  ``The anti-de Sitter Gott universe: A rotating BTZ wormhole,''
  Class.\ Quant.\ Grav.\  {\bf 16}, 3095 (1999)
  [arXiv:gr-qc/9905030].


\bibitem{as}
P.~C. Aichelburg and R.~U. Sexl,
``On the gravitational field of a massless particle,''
J. Gen.\ Rel.\ Grav.\ {\bf 2}, 303 (1971).

\bibitem{dh}
  T.~Dray and G.~'t Hooft,
  ``The Gravitational Shock Wave Of A Massless Particle,''
  Nucl.\ Phys.\  B {\bf 253}, 173 (1985).


\bibitem{hottatanaka}
  M.~Hotta and M.~Tanaka,
  ``Shock wave geometry with nonvanishing cosmological constant,''
  Class.\ Quant.\ Grav.\  {\bf 10}, 307 (1993).

\bibitem{podolskygriffiths}
  J.~Podolsky and J.~B.~Griffiths,
  ``Impulsive gravitational waves generated by null particles in de Sitter and
  anti-de Sitter backgrounds,''
  Phys.\ Rev.\  D {\bf 56}, 4756 (1997).


\bibitem{horoitzhaki}
  G.~T.~Horowitz and N.~Itzhaki,
  ``Black holes, shock waves, and causality in the AdS/CFT correspondence,''
  JHEP {\bf 9902}, 010 (1999)
  [arXiv:hep-th/9901012].

\bibitem{kaloperterning}
  N.~Kaloper and J.~Terning,
  ``How black holes form in high energy collisions,''
  Gen.\ Rel.\ Grav.\  {\bf 39}, 1525 (2007)
  [arXiv:0705.0408 [hep-th]].

\bibitem{death1}
  P.~D.~D'Eath,
  ``High Speed Black Hole Encounters And Gravitational Radiation,''
  Phys.\ Rev.\  D {\bf 18}, 990 (1978);

\bibitem{death2}
P.~D.~D'Eath and P.~N.~Payne,
  ``Gravitational Radiation In High Speed Black Hole Collisions. 1.
  Perturbation Treatment Of The Axisymmetric Speed Of Light Collision,''
  Phys.\ Rev.\  D {\bf 46}, 658 (1992).

\bibitem{ss}
  K.~Skenderis and S.~N.~Solodukhin,
  ``Quantum effective action from the AdS/CFT correspondence,''
  Phys.\ Lett.\  B {\bf 472}, 316 (2000)
  [arXiv:hep-th/9910023].

\bibitem{skenderis}
  S.~de Haro, S.~N.~Solodukhin and K.~Skenderis,
  ``Holographic reconstruction of spacetime and renormalization
  in the  AdS/CFT correspondence,''
  Commun.\ Math.\ Phys.\  {\bf 217}, 595 (2001)
  [arXiv:hep-th/0002230].

\bibitem{bjp}
  A.~Bialas, R.~A.~Janik and R.~Peschanski,
  ``Unified description of Bjorken and Landau 1+1 hydrodynamics,''
  Phys.\ Rev.\  C {\bf 76}, 054901 (2007)
  [arXiv:0706.2108 [nucl-th]].

\bibitem{henneaux}
  M.~Banados, M.~Henneaux, C.~Teitelboim and J.~Zanelli,
  ``Geometry of the (2+1) black hole,''
  Phys.\ Rev.\  D {\bf 48}, 1506 (1993)
  [arXiv:gr-qc/9302012].

\bibitem{carlip}
  S.~Carlip,
  ``The (2+1)-Dimensional black hole,''
  Class.\ Quant.\ Grav.\  {\bf 12}, 2853 (1995)
  [arXiv:gr-qc/9506079].

\bibitem{Bjorken:1982qr}
  J.~D.~Bjorken,
  ``Highly Relativistic Nucleus-Nucleus Collisions: The Central Rapidity
  Region,''
  Phys.\ Rev.\  D {\bf 27}, 140 (1983).

\bibitem{Balasubramanian:2003kq}
  V.~Balasubramanian, A.~Naqvi and J.~Simon,
  ``A multi-boundary AdS orbifold and DLCQ holography: A universal  holographic
  description of extremal black hole horizons,''
  JHEP {\bf 0408}, 023 (2004)
  [arXiv:hep-th/0311237].

\bibitem{lappi} See, for example,
  T.~Lappi and L.~McLerran,
  ``Some features of the glasma,''
  Nucl.\ Phys.\  A {\bf 772}, 200 (2006)
  [arXiv:hep-ph/0602189].

\bibitem{malda-stro}
J.~Maldacena and A.~Strominger,
``AdS${}_3$ Black Holes and a
Stringy Exclusion Principle'',
JHEP {\bf 9812}, 005 (1998)
[arXiv:hep-th/9804085].

\bibitem{lappigsqmu}
  T.~Lappi,
  ``Wilson line correlator in the MV model: relating the glasma to deep
  inelastic scattering,''
  arXiv:0711.3039 [hep-ph].

\bibitem{lappiendens}
  T.~Lappi,
  ``Energy density of the glasma,''
  Phys.\ Lett.\  B {\bf 643}, 11 (2006)
  [arXiv:hep-ph/0606207].

\bibitem{landau}
L.~D.~Landau, Izv.\ Akd.\ Naur.\ Ser.\ Fiz.\ {\bf 17}, 51 (1953).



\end{thebibliography}
\end{document}